\documentclass[11pt,twoside]{article}
\usepackage{asp2010}

\resetcounters

\begin{document}

\title{IC4663: the first unambiguous [WN] Wolf-Rayet central star of a planetary nebula}
\author{Brent Miszalski,$^{1,2}$ Paul A. Crowther,$^3$ Orsola De Marco,$^4$ Joachim K\"oppen,$^5$ Anthony F.J. Moffat,$^{6}$ Agn\`es Acker$^5$ and Todd C. Hillwig$^7$
\affil{$^1$South African Astronomical Observatory, PO Box 9, Observatory, 7935, South Africa}
\affil{$^2$Southern African Large Telescope Foundation, PO Box 9, Observatory, 7935, South Africa}
\affil{$^3$Department of Physics and Astronomy, Hounsfield Road, University of Sheffield, Sheffield S3 7RH, UK}
\affil{$^4$Department of Physics and Astronomy, Macquarie University, Sydney, NSW 2109, Australia}
\affil{$^5$Observatoire astronomique de Strasbourg, Universit\'e de Strasbourg, CNRS, UMR 7550, 11 rue de l'Universit\'e, F-67000 Strasbourg, France}
\affil{$^6$D\'ept. de physique, Univ. de Montr\'eal C.P. 6128, Succ. Centre-Ville, Montr\'eal, QC H3C 3J7, and Centre de recherche\\ en astrophysique du Qu\'ebec, Canada}
\affil{$^7$Department of Physics and Astronomy, Valparaiso University, Valparaiso, IN 46383, USA}}

\begin{abstract}
   Several [WC]-type central stars of planetary nebulae (PNe) are known to mimic the spectroscopic appearance of massive carbon-rich or WC-type Wolf-Rayet stars. In stark contrast, no [WN]-type central stars have yet been identified as clear-cut analogues of the common nitrogen-rich or WN-type Wolf-Rayet stars. We have identified the [WN3] central star of IC4663 to be the first unambiguous example in PNe. The low luminosity nucleus and an asymptotic giant branch (AGB) halo surrounding the main nebula prove the bona-fide PN nature of IC4663. Model atmosphere analysis reveals the [WN3] star to have an exotic chemical composition of helium (95\%), hydrogen ($<$2\%), nitrogen (0.8\%), neon (0.2\%) and oxygen (0.05\%) by mass. Such an extreme helium-dominated composition cannot be predicted by current evolutionary scenarios for hydrogen deficient [WC]-type central stars. Only with the discovery of IC4663 and its unusual composition can we now connect [WN] central stars to the O(He) central stars in a second H-deficient and He-rich evolutionary sequence, [WN]$\to$O(He), that exists in parallel to the carbon-rich [WC]$\to$PG1159 sequence. This suggests a simpler mechanism, perhaps a binary merger, can better explain H-deficiency in PNe and potentially other H-deficient/He-rich stars. In this respect IC4663 is the best supported case for a possible merged binary central star of a PN. 
\end{abstract}
\section{Introduction}
The immediate progenitors of some non-DA white dwarfs (WDs) may be found in the H-deficient post-Asymptotic Giant Branch (AGB) nuclei of PNe. Most common amongst H-deficient nuclei are those with fast and dense stellar winds that mimic the carbon sequence of massive Wolf-Rayet (WR) stars \citep{2008ASPC..391...83C}. These stars, denoted as [WC] stars to separate them from their massive WC counterparts, have atmospheres rich in carbon, oxygen and helium as readily seen in their remarkable emission line spectra \citep{1998MNRAS.296..367C,2003A&A...403..659A}. Stellar atmosphere analyses have demonstrated a strong similarity between compositions of [WC] stars and PG1159 stars \citep{2006PASP..118..183W,2008ASPC..391...83C}, implying an evolutionary sequence [WC]$\to$PG1159$\to$WD. \citet{2006PASP..118..183W} give several examples of He-rich/H-deficient stars that do not seem to fit this sequence. Perhaps the least understood are the O(He) stars, hot post-AGB stars ($T_\mathrm{eff}\ge100$ kK) with atmospheres dominated by helium (\cite{1998A&A...338..651R}; Reindl et al. these proc.). The final destiny of O(He) stars will be either a helium-rich DO WD or a DA WD (if it contains some residual hydrogen), but essentially nothing is known about their origin. 

As two O(He) stars are surrounded by PNe, it would not be unexpected to find a progenitor amongst other central stars of PNe. No other central stars were known to have comparable He-dominated compositions, i.e. a helium mass fraction $\ga$90\%, until we studied the unique central star of IC4663 \citep{2012MNRAS.423..934M}. Unlike all other WR central stars, it is the first proven case of a central star mimicking the nitrogen sequence of massive WR stars, a true [WN] star! Here we summarise its properties from which we were compelled to propose a new, second He-rich/H-deficient post-AGB evolutionary sequence [WN]$\to$O(He)$\to$WD.

\subsection{First of a kind: IC4663}
While there are $\ga$100 [WC] central stars known \citep{2011MNRAS.414.2812D}, only a handful of [WN] candidates have been identified. These candidates have often turned out to be massive WN stars with ejecta nebulae \citep{2010MNRAS.409.1429S}, due to the very large uncertainty in estimating distances to PNe, or could not be proven one way or the other. The most promising candidates were LMC-N66, which has an uncomfortably high luminosity \citep{2003A&A...409..969H}, and PMR5 \citep{2003MNRAS.346..719M}, for which several indicators now suggest it is a reddened massive WN star \citep{2010AIPC.1273..219T}. \citet{2010A&A...515A..83T} studied the WR-like central star of PB8 (a bona-fide PN) and found an unusual atmospheric composition unlike [WC] stars. The spectral classification of PB8 is not entirely clear and may be a hybrid [WN/WC] type rather than a pure [WN] type. Another [WN] candidate is the nucleus of Abell~48 (Bojicic et al. these proc.), however a definitive study has yet to be published for this object.

\begin{figure}
   \begin{center}
      \includegraphics[scale=0.45,angle=270]{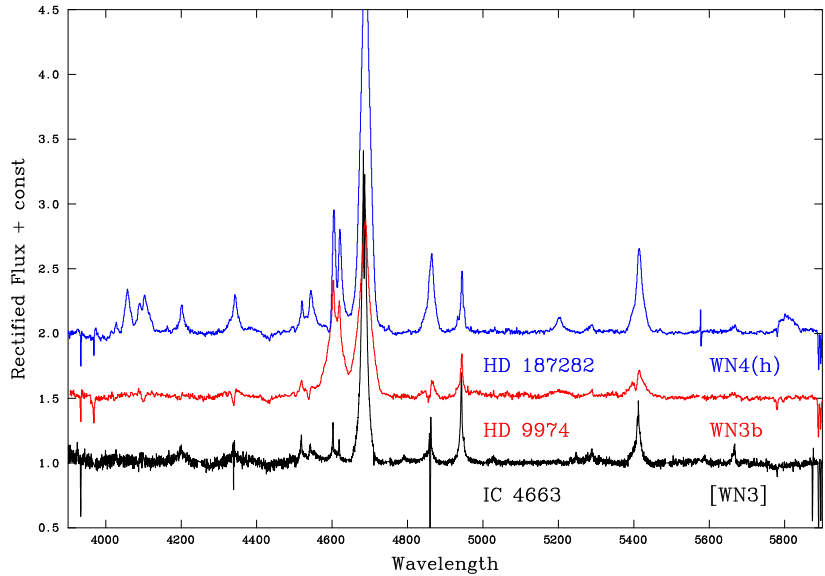}
   \end{center}
   \caption{Gemini GMOS spectrum of the [WN3] central star of IC4663 in comparison to massive WN stars of similar spectral type.}
   \label{fig:montage}
\end{figure}

\begin{figure}
   \begin{center}
      \includegraphics[scale=0.5]{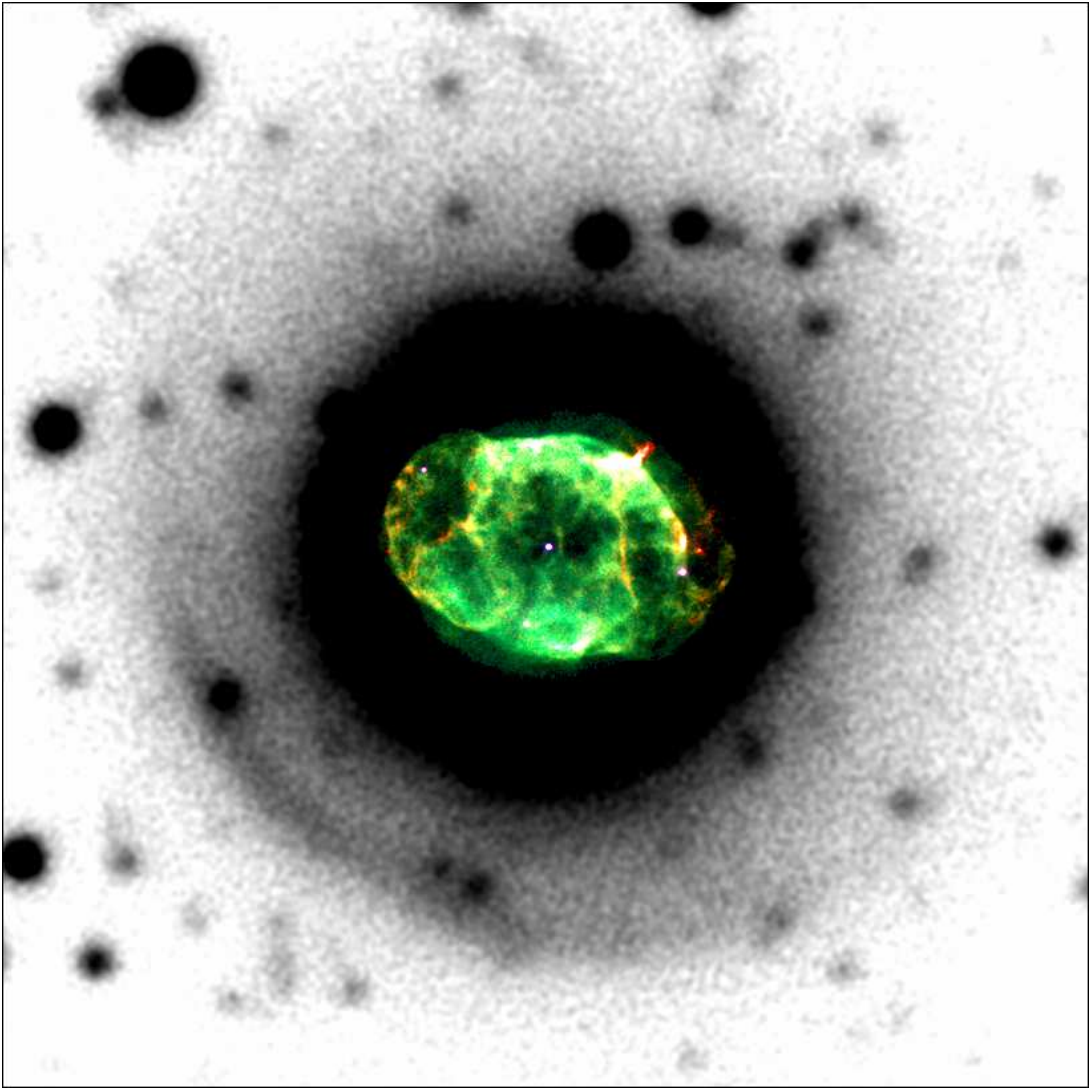}
   \end{center}
   \caption{A newly discovered faint AGB halo (GMOS [OIII] image, background) surrounds an \emph{HST} colour-composite image of IC4663 where $F658N$ is red, $F502N$ is green and $F555W$ is blue \citep{2012MNRAS.423..934M}. AGB haloes are a telltale feature of PNe \citep{2003MNRAS.340..417C}. The image dimensions are $60\times60$ arcsec$^2$ with North up and East to the left.}
   \label{fig:neb}
\end{figure}

We obtained imaging and spectroscopy of IC4663 (PN G346.2$-$08.2) in June and July 2011 with GMOS on Gemini South \citep{2004PASP..116..425H}. Figure \ref{fig:montage} shows our spectrum of the previously unstudied central star which has a [WN3] spectral type following \citet{1996MNRAS.281..163S}. The luminosity of the $V=16.9$ mag central star is always consistent with a PN, i.e. 4--6 mag fainter than massive WN stars of the same spectral type for all reasonable distances \citep{2006A&A...457.1015H}. At an assumed distance of 3.5 kpc \citep{2008ApJ...689..194S} this corresponds to $L=4000$ $L_\odot$ and $M_V=+3.1$ mag. Furthermore, the nebula is clearly a PN with an elliptical morphology, a low expansion velocity of 30 km s$^{-1}$ \citep{2007ApJS..169..289H} and a newly discovered AGB halo (Fig. \ref{fig:neb}). A chemical abundance analysis of the nebular emission lines reveals an approximately solar abundance pattern with slight enhancements in helium, nitrogen and neon, and carbon may be underabundant.

\section{Evolutionary status of IC4663}
We analysed the GMOS spectrum with the CMFGEN NLTE model atmosphere code \citep{1998ApJ...496..407H}, the full details of which may be found in \citet{2012MNRAS.423..934M}. The physical properties are not dissimilar from hot [WO] central stars, namely $T_*=140\pm20$ kK, $v_\infty=1900$ km s$^{-1}$, log $g=6.1$ (assuming $M=0.6$ $M_\odot$), and for $d=3.5$ kpc we find $R_*=0.11$ $R_\odot$, $\dot{M}=1.8\times10^{-8}$ $M_\odot$ yr$^{-1}$, and $L=4000$ $L_\odot$. The most surprising results from this analysis is the atmospheric composition which contains by mass helium (95\%), hydrogen ($<2$\%), carbon ($<0.1$\%), nitrogen (0.8\%), oxygen (0.02\%) and neon (0.2\%). Clearly the He-dominated composition of IC4663 does not fit the [WC]$\to$PG1159 sequence \citep{2006PASP..118..183W}, but rather matches closely the composition of O(He) stars \citep{1998A&A...338..651R}. This strongly suggests there is a second, parallel H-deficient and He-rich post-AGB evolutionary sequence [WN]$\to$O(He). There has been some suggestion that this sequence exists \citep{2012IAUS..283..196W}, however only with our study of IC4663 has this now become tangible. The O(He) stars are therefore the He-rich analogues of PG1159 stars.  The nebular evolutionary status of IC4663 and other PNe with H-deficient central stars are also consistent with this evolutionary sequence \citep{2012MNRAS.423..934M}.

None of the scenarios proposed to explain the composition of [WC] or PG1159 stars can produce the composition of IC4663 \citep{2006PASP..118..183W}, suggesting there is another simpler mechanism that may be responsible for creating H-deficient central stars of PNe. The most promising of which may be a double-degenerate merger \citep{2002MNRAS.333..121S}. Reindl et al. (these proc.) showed that the composition of IC4663 agreed well with predictions of the slow merger of two He-WDs described by \citet{2012MNRAS.419..452Z}. This picture is consistent with the lack of radial velocity variability seen in our three GMOS spectra and makes IC4663 the best supported case for a possible merged binary central star of a PN. 

\section{Implications for central star classification}
How does IC4663 affect existing classification schemes for PNe central stars? Firstly, the discovery of its [WN3] central star means that there should exist other [WN] stars that mimic WN stars of later sub-types. We have identified several late-[WN] stars that we are analysing in the context of establishing the basic characteristics of the [WN] sequence in PNe. 

\begin{figure}
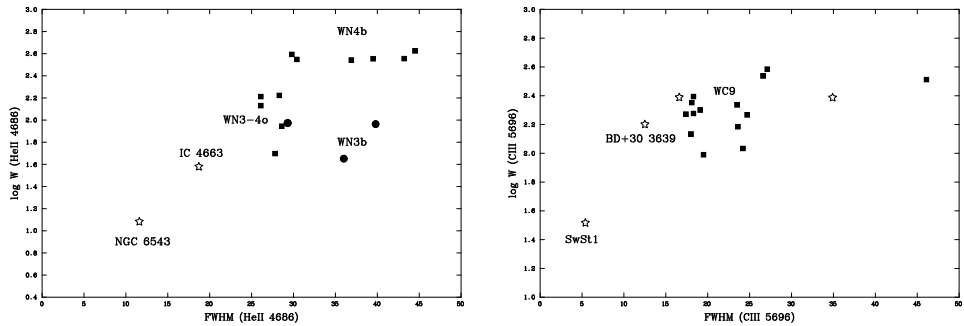

   \begin{center}
      \includegraphics[scale=0.25,angle=270]{miszalski_fig3a.eps}
      \includegraphics[scale=0.25,angle=270]{miszalski_fig3b.eps}
   \end{center}
   \caption{\emph{(left)} The position of IC4663 ([WN3]) in log $W$/FWHM space of He~II $\lambda$4686 with NGC6543 (Of/WN) and massive WN3/WN4 stars (filled symbols). \emph{(right)} A similar plot for BD+30~3639 ([WC9]), SwSt~1 (Of/WC) and massive WC9 stars using C~III $\lambda$5696.} 
   \label{fig:widths}
\end{figure}

Secondly, there is a close relationship between massive Of and WN stars \citep{2011MNRAS.416.1311C}. Intermediate Of/WN types exist and in PNe these have sometimes been classified as Of-WR central stars \citep{1991IAUS..145..375M}. The best example of which is NGC6543 which exhibits a WN-like spectrum and an H-rich atmosphere \citep{M90,G08}. A useful diagnostic plot for distinguishing bona-fide [WN] stars from Of or Of/[WN] types may be that found in Fig. \ref{fig:widths}. The equivalent width $W$ of HeII $\lambda$4686 for IC4663 is comparable to WN stars of similar type, whereas the full width at half-maximum (FWHM) is lower. Both values are lower in the isolated NGC6543 of Of/WN type. Figure \ref{fig:widths} also shows that BD+30~3639 ([WC9]) follows a similar pattern c.f. WC9 stars in the CIII $\lambda$5696 emission line and that SwSt~1 is a carbon-rich analogue of NGC6543. Note that BD+30~3639 has always been considered a [WC9] star. \citet{2012MNRAS.423..934M} also demonstrated that the transformed radius \citep{2006A&A...457.1015H} for IC4663 is comparable to WN stars of similar spectral types, ruling out any Of or Of/WN classification of the [WN3] star. Other examples of Of or Of/WN types may have mistakenly been classified as so-called `weak emission line' central stars, but this usage should be discontinued in favour of a new scheme that is cognisant of Of/WN types and close binary central stars (e.g. \cite{2011A&A...531A.158M}).

\bibliography{miszalski}
\end{document}